# Performance Evaluation of LLMs in Automated RDF Knowledge Graph Generation

Ioana Ramona Martin[1], Tudor Cioara[1*], Ionut Anghel[1] and Gabriel Arcas[2]

[1]Distributed Systems Research Laboratory, Computer Science Department, Technical University of Cluj-Napoca, G. Barițiu 26-28, 400027 Cluj-Napoca, Romania; martin.io.ioana@student.utcluj.ro; tudor.cioara@cs.utcluj.ro; ionut.anghel@cs.utcluj.ro

[2]Department of Cloud & DevOpsRebelDot, Buftea 1, 400606 Cluj-Napoca, România; gabriel.arcas@rebeldot.com

*Corresponding author

**Abstract:** Cloud systems generate large, heterogeneous log data containing critical infrastructure, application, and security information. Transforming these logs into RDF triples enables their integration into knowledge graphs, improving interpretability, root-cause analysis, and cross-service reasoning beyond what raw logs allow. Large Language Models (LLMs) offer a promising approach to automate RDF knowledge graph generation; however, their effectiveness on complex cloud logs remains largely unexplored. In this paper, we evaluate multiple LLM architectures and prompting strategies for automated RDF extraction using a controlled framework with two pipelines for systematically processing semi-structured log data. The extraction pipeline integrates multiple LLMs to identify relevant entities and relationships, automatically generating subject-predicate-object triples. These outputs are evaluated using a dedicated validation pipeline with both syntactic and semantic metrics to assess accuracy, completeness, and quality. Due to the lack of public ground-truth datasets, we created a reference Log-to-KG dataset from OpenStack logs using manual annotation and ontology-driven methods, enabling objective baseline. Our analysis shows that Few-Shot learning is the most effective strategy, with Llama achieving a 99.35% F1 score and 100% valid RDF output while Qwen, NuExtract, and Gemma also perform well under Few-Shot prompting, with Chain-of-Thought approaches maintaining similar accuracy. One-Shot prompting offers a lighter but effective alternative, while Zero-Shot and advanced strategies such as Tree-of-Thought, Self-Critique, and Generate-Multiple perform substantially worse. These results highlight the importance of contextual examples and prompt design for accurate RDF extraction and reveal model-specific limitations across LLM architectures.

**Keywords:** Artificial Intelligence, Large Language Models, Prompting Techniques, RDF, Knowledge Graphs, Reference dataset

## 1. Introduction

Cloud applications are producing massive amounts of unstructured logs during their lifecycles. They are highly useful for debugging or ensuring the well-being of a system, but often, they imply laborious work to be understood and correlated [1]. These logs contain rich semantic information about system events, failures, dependencies, and configurations. However, much of the content is unstructured or varies in format and terminology depending on the source, operation, or event type. This variability challenges the use of deterministic parsing methods and makes reliable information extraction difficult [2]. As a result, extracting meaningful relationships requires understanding context, which rule-based or regex-based approaches cannot capture in a reliable manner [3]. Therefore, transforming cloud logs into structured data in the form of RDF triples enables their integration into a knowledge graph (KG), significantly improving interpretability, root-cause investigation, and cross-service reasoning, which are difficult with raw logs alone [4]. Structuring logs into a knowledge graph provides the first step for transforming raw data into actionable insights, and moreover, when combined with GraphRAG [5], enables Agentic AI systems capable of autonomously analysing, reasoning over, and acting on the extracted knowledge.

Recent advances in large language models (LLMs) have improved knowledge graph (KG) construction and querying, enabling automated extraction of entities, relationships, and knowledge completion across diverse domains. Unlike deterministic parsing methods, LLMs can handle semi-structured and free-text data, interpreting contextual nuances and variations in message content [6]. Their ability to generalize across heterogeneous log sources allows for a more reliable and flexible extraction strategy, accommodating different types of logs, events or applications. LLM-based frameworks [7, 8] demonstrate cost-effective and high-precision triplet extraction, while integration with ontologies and neuro-symbolic techniques enhances reasoning capabilities, standardises outputs, and reduces human effort [9]. Prompting and fine-tuning strategies further improve the quality of extracted knowledge [10], and domain-specific applications illustrate feasibility in heterogeneous datasets [11, 12]. LLMs have also been applied to querying and question answering over KGs, supporting complex SPARQL queries, multi-document reasoning, and low-resource scenarios [13, 14].

However, despite these achievements, key challenges remain. Distributed application logs present highly heterogeneous, noisy, and domain-specific data that differ substantially from the curated datasets commonly used in prior works [15]. Most studies have evaluated single models or isolated techniques, leaving the comparative performance of different LLMs and fine-tuning strategies largely unexplored [16]. Issues such as hallucination, noisy extraction, scalability, and the need for automated verification further complicated the KG construction in a real-world distributed environment [17]. All these highlight the need for a systematic evaluation of multiple LLMs and techniques for constructing KGs from distributed logs. Such a study can reveal how different models handle heterogeneous and noisy data, the relative effectiveness of prompting versus fine-tuning strategies, and the trade-offs in precision and completeness, providing practical guidance for LLMs adoption for log processing tasks.

In this paper, we aim to evaluate multiple LLM architectures and fine-tuning techniques in terms of their effectiveness and efficiency for automated structured RDF KG knowledge extraction out of cloud systems logs. To support systematic and comparative evaluation, we designed a framework featuring two pipelines that establish a controlled testbed in which selected LLMs are applied to heterogeneous and semi-structured log data. This setup facilitates the analysis of model performance across different prompt engineering strategies and highlights the strengths and limitations of various approaches. Detailed implementation aspects, including the employed models, technology stack, and supporting tools, are presented to ensure reproducibility and transparency.

The LLM-based RDF extraction pipeline transforms raw, unstructured cloud logs into structured RDF triples, which serve as the foundational components of an RDF-based knowledge graph. The pipeline integrates various LLMs and prompting techniques to extract relevant information from log messages, applies semantic mappings, and automatically generates subject-predicate-object triples in RDF. These triples are subsequently integrated into a KG, enabling a semantic representation of cloud logs and supporting expressive queries that uncover insights not readily obtainable from raw log data.

The LLM-generated outputs are passed to a dedicated validation and evaluation pipeline, where multiple metrics are computed to assess the accuracy, completeness, and overall quality of the extracted RDF triples. Objective evaluation requires a reference dataset at the individual log-entry level since the LLM outputs can be of a non-deterministic nature, and there is a risk of incorrect yet plausible triples. As no public ground-truth datasets exist for LLM-based KG extraction, we constructed a dedicated Log-to-KG reference dataset using OpenStack cloud logs from a public repository available on GitHub [18]. The reference dataset was constructed through a combination of manual annotation and ontology-driven methods, enabling objective benchmarking of the generated triples. The evaluation pipeline supports two complementary perspectives: strict syntactic correctness and semantic extraction quality. Syntactic evaluation requires exact matching of RDF's subjects, predicates, and objects against the reference Log-to-KG dataset, measuring schema compliance. In contrast, semantic evaluation allows minor formatting and normalisation variations, considering a triple valid if the semantic relation is correctly extracted.

Our experiments evaluated the performance of various LLMs in extracting RDF data from unstructured log files, comparing Zero-Shot, One-Shot, Few-Shot, and advanced prompting techniques. Few-Shot learning was the most effective strategy, achieving the highest performance across nearly all models. Llama, with few-shot prompting, achieved very good results, with a 99.35% F1 score and 100% valid RDF output. Other models, including Qwen (99.22% F1), NuExtract (98.47% F1), and Gemma (94.81% F1), also performed well using few-shot approaches. The few-shot variant combined with Chain-of-Thought prompting maintained similarly high performance, with Llama achieving a 99.32% F1 score and perfect RDF validity. One-shot prompting proved to be a lighter but effective alternative, with multiple models achieving over 85% F1 scores using both One-shot and One-shot with Chain-of-Thought prompting. These results demonstrate that LLMs are highly capable of extracting structured RDF information when provided with contextual examples. In contrast, Zero-Shot approaches failed most of the time, with most models producing F1 scores below 30% (semantic) and 20% (syntactic). Qwen Zero-Shot managed only 0.6% valid RDF outputs, while DeepSeek achieved 18.7% validity with an F1 score of 9.15%. Advanced prompting techniques also performed worse relative to Few-Shot baselines: Tree-of-Thought reached a best F1 of 59.02%, Self-Critique 64.38%, and Generate-Multiple 68.20%. DeepSeek was consistently struggling across all techniques, never exceeding 91.60% semantic F1 or 51.80% syntactic F1, and producing high rates of invalid or regex-fallback outputs even with advanced prompting. Overall, the results highlight that prompt design and contextual examples are the most important elements for accurate RDF extraction. Model-specific limitations, as seen in DeepSeek, further underscore the variability in LLM performance across different architectures and prompting strategies.

The rest of the paper is organised as follows. Section 2 reviews related work on LLM-based log processing and knowledge graph extraction. Section 3 presents the proposed LLM-based RDF extraction approach, focusing on model selection and prompting techniques. Section 4 describes the knowledge graph modelling and the construction of the Log-to-KG reference dataset. Section 5 details the implementation aspects, including the employed models, technology stack, and supporting tools. Section 6 reports the evaluation results for both syntactic and semantic assessments across various LLMs and prompting strategies. Section 7 discusses the obtained results, with particular emphasis on predicate extraction and prompting performance. Finally, Section 8 concludes the paper and presents the future work.

## 2. Related work

Recent advances in large language models have transformed modelling, automatic construction, and enrichment of knowledge graphs, enabling more scalable, semantically rich knowledge representations.

**Several works focus on leveraging LLMs as central agents for KG construction**. In [19], a unified LLM agent is proposed for two main tasks: Relational Triplet Extraction (RTE) and Knowledge Graph Completion (KGC). The RTE step extracts triplets from urban text data, while KGC predicts relationships between existing entities. This framework proved more cost-effective than comparable tools, including GPT-4. Similarly, [20] addresses limitations of traditional methods that struggle with structured data, employing multiple LLM agents for text extraction, segmentation, triplet extraction, and knowledge integration. Evaluations demonstrated higher precision than state-of-the-art approaches, especially due to improved text comprehension. Generate-on-Graph [21] also combines LLMs and KGs to leverage shared knowledge. The proposed training-free Generate-on-Graph (GoG) method introduces an intuitive Thinking–Searching–Generating framework that allows LLMs to generate missing facts during reasoning. Experimental results on two datasets show consistent improvements over existing methods. KGLM [22] framework further extends this idea in enterprise and biomedical contexts, integrating LLMs with enterprise KGs (EKGs) and ontological reasoning to enhance explainability, natural language interaction, and domain-specific responses. The approach uses a neuro-symbolic pipeline that connects LLMs with CHASE-based reasoning systems, such as Vadalog, to support explainable natural language interfaces over enterprise knowledge graphs.

**LLMs can be integrated with ontologies to improve KG construction**. Current studies explore multiple strategies for unifying LLMs and KGs: KG-enhanced LLMs improve knowledge understanding during preprocessing and inference, LLM-augmented KGs leverage LLMs for embedding, completion, graph-to-text generation, and question answering, and synergistic approaches use both components equally to enhance reasoning capabilities [23]. In [24], a semi-automated approach reduces human effort in ontology definition, enabling LLMs to formulate competency questions, develop ontologies, and construct KGs, while humans remain in the loop for evaluation and adjustments. The authors use of a concrete case study on deep learning literature and include a judge LLM for evaluating RAG outputs and extracted KG concepts. Further, embedding Chain-of-Thought (CoT) reasoning into LLMs, guided by ontologies, improves triplet standardization and extraction quality [25]. Ontologies provide domain-specific knowledge and guide the LLM during data processing, while zero-shot evaluation shows consistent accuracy and reduced variability in relationship expression. Ontology-based verification of RDF triples also enhances extraction accuracy in pipelines that include LLMs [26]. A two-stage pipeline is proposed, consisting of initial entity candidate extraction followed by refinement and alignment with canonical knowledge graph entities and relations, with final triplet validation using Wikidata relation constraints.

**Prompting strategies have become a key method for guiding LLMs in structured data extraction**. Zero-shot prompting extracts structured outputs without labelled examples, while in-context learning addresses output variability by providing guiding schemas [27]. Few-shot prompting further enhances performance by supplying examples within prompts, demonstrating that both model size and the quality of the underlying knowledge base significantly affect triple extraction [28]. Multi-prompt strategies, including negative examples, have been applied to context-dependent texts such as Regional Trade Agreements, improving performance and generating meaningful RDF triples [29]. Fine-tuning and model adaptation techniques also improve LLM performance. For example, data-augmented fine-tuning enhances triple quality compared to off-the-shelf models like GPT-4 [30]. Task-oriented dialogue systems employ prompts incorporating ontologies, input text, and task descriptions, showing that well-structured prompts lead to higher-quality triples [31]. Combining LLMs with KG embeddings improves generalisation and reasoning on unseen data [32], while property extraction in scholarly KGs demonstrates that fine-tuning and advanced prompt engineering can overcome inconsistencies [33].

**LLMs have been applied in various domains to construct KGs and extract RDF triples**. DeepSeek LLM has been used in LLM-based knowledge graph construction framework for task-oriented semantic communication, covering corpus collection, entity and relation extraction, knowledge base generation, and dynamic updates, enhanced through prompt engineering and few-shot learning [34]. In medicine, LLMs enhance reasoning, scalability, and heterogeneity while improving precision in KG construction [35]. Framework Materials [36] leverage LLMs to extract properties, relationships, entities, and applications for automated KG pipelines, supporting data retrieval, mining, and question answering. Experiments on the WebNGL dataset [37] demonstrate improved semantic and structural accuracy in triple extraction using cosine similarity and graph edit distance. Other works focus on semantic annotation and evaluation of triples. LLM-based Named Entity Recognition (NER) outperforms traditional and fine-tuned models, and chain-of-thought prompting further enhances information retrieval [38]. Finally, [39] evaluates 12 LLMs against human judgment for triple validation. Although LLMs improve completeness assessments, traditional methods still outperform them in some cases. Additional strategies address LLM limitations such as noise, hallucination, and domain-specific knowledge extraction. Techniques include Entity-Centric

Iterative Text Denoising, Knowledge-Aware Instruction Tuning [40], SPARQL-based refinement of triples [41], and multi-stage pipelines combining extraction, learning, and evaluation [42]. KG completion using LLMs, where extracted triples are used as prompts for completion, achieves state-of-the-art performance [43].

Beyond KG construction, **LLMs are increasingly used for querying, reasoning, and question answering over KGs**. In [44], a multi-agent framework is proposed to translate natural language into graph query language, with three agents: Preprocessor (data processing), Generator (query generation), and Refiner (improving queries based on execution results). This approach outperformed benchmarks for context-aware graph queries. For semantic query processing, [45] decomposes complex queries into simple interconnected triples, which are solved individually and recombined. This method allows LLMs to produce accurate SPARQL results efficiently, surpassing traditional information retrieval methods in both accuracy and time. In generative graph analytics, [46] demonstrates that LLMs can identify non-obvious similarities across predicates in different KGs, support multi-document question answering via KG prompting, and maintain contextual coherence for complex queries. Effective prompt design is critical to ensure accurate interpretation of graph structures. Finally, [47] presents a fine-tuned LLM for low-resource KG question answering using Generative Adversarial Imitation Learning (GAIL). This approach reduces distribution shift between synthetic and real queries and leverages automated SPARQL-to-natural language question generation, improving QA performance and adaptability in real-world KG applications.

## 3. LLM-Based RDF extraction

Our solution allows for the selection and integration of different types of LLMs depending on task-specific needs. Therefore, in this section, we first define the criteria used for evaluating candidate LLMs, reflecting and justifying the set of LLMs incorporated into the framework. Following this, we describe the prompt engineering techniques employed and the common structure shared across all prompts.

### 3.1. LLM models selection

The models were selected to perform the core task of semantic extraction from unstructured or semi-structured OpenStack logs. The selection criteria used by us refer to the models' pre-training and open-source features and to the capabilities of following natural language instructions to generate structured outputs. They also needed to be robust with technical and semi-structured log data, lightweight and comparable in scale (4B–8B parameters) to ensure fair evaluation, and include diversity in training objectives, such as general-purpose instruction-tuned, extraction-oriented, and distilled models. Finally, practical deployment and efficiency were essential, requiring compatibility with the Transformers framework to support multiple experiments with different prompting strategies.

Six language models were selected for this project according to the criteria outlined previously. Table 1 summarizes the key characteristics of the selected models, their strengths, limitations, and suitability for the task. Each model possesses distinct features that may influence extraction performance, robustness, and evaluation results. All chosen models are publicly available on Hugging Face and free to use.

**Table 1. Selected LLM models for framework integration**

| Model | Parameters | Key Advantages | Limitations | Suitability for Log-to-RDF KG Construction |
|---|---|---|---|---|
| Qwen2.5-7B-Instruct [48] | 7B | Multilingual, instruction-tuned, produces structured output, robust on semi-structured text | Output sensitive to prompt phrasing; variability in consistency | Good for structured RDF generation from semi-structured logs |
| NuExtract-2.0-8B [49] | 8B | Fine-tuned for information extraction, handles large documents, concise outputs | Less flexible, may underperform with broad context | Effective for precise extraction from unstructured/semi-structured logs |
| Mistral-7B-Instruct-v0.2 [50] | 7B | Balanced efficiency and reasoning, moderate computational requirements | May miss subtle entities/relationships; struggles with niche jargon | Suitable for general log extraction with limited resources |
| Llama-3.1-8B-Instruct [51] | 8B | Large context window, widely adopted baseline, strong reasoning | Higher computational cost; risk of over-generation | Good for long prompts and diverse log scenarios |
| gemma-3-4b-it [52] | 4B | Lightweight, clear and logically consistent output, efficient | Reduced capacity for complex reasoning; limited long-context handling | Suitable for smaller-scale or quick extraction tasks |
| DeepSeek-R1-Distill-Llama-8B [53] | 8B | Structured output, efficient reasoning, strong accuracy-efficiency trade-off | Sensitive to complex prompt engineering | Effective for RDF extraction with structured reasoning, careful prompting required |

We have included models ranging from lightweight architectures like gemma-3-4b-it to larger, reasoning-focused models such as DeepSeek-R1-Distill-Llama-8B, to capture a wide spectrum of computational trade-offs, context handling, and output consistency. Each model presents distinct advantages and limitations, for example, some excel at concise extraction (NuExtract-2.0-8B), while others handle long context windows (Llama-3.1-8B-Instruct) or complex reasoning (DeepSeek-R1). This diversity allows for a systematic comparison of LLM performance, and efficiency, providing insights into which architecture and prompting strategies are best suited for automated KG construction from noisy, heterogeneous logs.

### 3.2. Prompt structure definition

We defined a structured prompt framework to guide the creation of actual prompts, ensuring that LLMs can transform semi-structured or unstructured input into structured, deterministic output. Through prompts the model was instructed to focus solely on semantic extraction rather than providing explanations ensuring outputs could be directly compared with the reference Log-to-KG dataset (see Section 4.2). Also, they were designed to reduce ambiguity, containing clear and precise instructions so that the model would only extract properties that exist. To minimize hallucinated entities or relationships, some prompts included an enforced schema, examples, or definitions to guide the model. All instructions and examples were aligned with the RDF schema for the KG, ensuring consistency and adherence to the defined structure.

Although each prompt is tailored for a specific prompting strategy, they all consist of common components, allowing for systematic construction, easy modification, and consistent application across different types of LLMs (see Figure 1). One of the common parts is the *[Task instruction]* where we insert a short and concise explanation of the task that the LLM needs to perform. Basically, it includes the information that the model needs to take the log line that is provided and then convert it into an RDF triple output by extracting the necessary subjects, predicates, and objects. A second common component is the *[Output Format Specification]*. The LLM is strictly told that the output needs to be expressed in TURTLE or TTL format to maintain the structure and to make it easier when comparing the results. Another common component refers to the *[Input Log Isolation]*. In each prompt, a single log line is provided, and the model is being told when that happens (usually at the end of the prompt). Finaly the prompt constraints and assumptions component are also common. In essence, this is telling the model not to provide explanation, commentary, or prose and only to output the result.

```
[Task Instruction]
You are tasked with converting a single OpenStack cloud log entry into an RDF triple. Extract the relevant subject, predicate, and object from the log line.

[Output Format Specification]
The output must be strictly in TURTLE (TTL) format. Do not include any explanation, commentary, or additional text. Only output the RDF triple.

[Constraints and Assumptions]
Only process the provided log line. Do not infer values from examples or explanations.
Do not generate explanations, prose, or commentary.
Ensure the RDF triple conforms to the RDF schema defined for this project.

[Optional Example Section] (if using few-shot or one-shot prompting)
Example log line: <sample log line>
Example RDF triple: <corresponding TTL triple>

[Input Log Isolation]
Here is the log line to process:
<Insert single log line here>
```

Figure 1. Common prompt components

The prompt design was an iterative process, shaped by the errors observed during model experiments. Initial prompts were simple, providing minimal instructions, which often led to inconsistent outputs, schema drifts, incomplete extractions, or the inclusion of explanations alongside the RDF triples. In other words, only telling the model to perform semantic extraction was not enough. Analysis of multiple LLM model outputs revealed recurring issues, including the generation of predicates not defined in the RDF schema, hallucinated entities or relationships, omission of relevant attributes, inconsistent formatting, invalid TTL structure, and the inclusion of verbose natural language. To address these issues the refinement strategy included explicitly restricting the model to a set of allowed predicates and entity types when the prompt technique permitted, tightening the output format specification, reinforcing to the model that it should only include the TTL output, not the free text explanation, and clarifying the required and optional nature of the properties.

### 3.3. Prompting strategies

Ten distinct prompting strategies were designed for the specific tasks of KG construction and evaluated, each of them having unique structure, different guidance levels and specific output constraints.

*Zero-Shot prompting (ZSP)* technique is the simplest of them all and the shortest in terms of text length. Its main goal is to only provide the instructions to the model, more specifically the task of receiving a cloud log as entry and then to extract the RDF triples and output the result in a specific format (see Figure 2).

For this kind of strategy, it is extremely important not to provide the model with any examples, since it should only rely on the prior training and its ability to interpret the given instructions. The usage of this technique is mostly relevant when it comes to what kind of properties and relationships the model can extract without having knowledge regarding the schema and regarding the vocabulary that is permitted for creating the RDF output.

> [Task Instruction]
> You are an expert in semantic web technologies and log-to-RDF transformation. Your task is to convert a single OpenStack log line into RDF triples by extracting the relevant subject, predicate, and object.

Figure 2. Zero-Shot prompt updates

*One-shot prompting (OSP)* provides the model with task instructions and a single example before the actual input is presented. The example in this case is a pair made of one log line and one RDF triple result, from the reference Log-to-KG dataset. In this way the models are understanding the RDF structure and the TTL output that it needs produce. Moreover, it is important for the model to have a demonstration of the predicate naming, the formatting, and, overall, the behaviour it needs to have during the processing. Furthermore, the example provided needs to contain many predicates. In this case, from the numerous examples that could have been given of multiple types of logs, we have decided that the most relevant one would be an HTTP request example, since it contains both resource objects and literals (see Figure 3). Finally, the prompt includes an additional constraint stating not to take any values from the example as we have observed that some LLMs might use the values from the example rather than processing the actual log line.

> [Constraints and Assumptions]
> Do not take any values from the example below; always process the actual log line provided.
>
> [Example Section]
> Example log line: <sample HTTP request log line>
> Example RDF triple: <corresponding TTL triple>

Figure 3. One-Shot prompt components addition and updates

*Few-Shots prompting (FSP)* technique is the longest and the richest. It keeps the full zero-shot prompt and adds the extra comment regarding the fact that the logs need to be processed, and the values should not be taken from the examples. As the name implies, few-shot prompting provides a small number of examples. Therefore, we have provided an example for each type of log covering API requests, resource management operations, file operation, and infrastructure-level events. This prompting technique provides the model with additional context regarding the extraction task, which can help improve consistency and alignment with the RDF schema illustrated in the examples. Potential advantages include reduced syntax errors, as the model is guided to produce valid TTL output, and improved schema adherence, since the examples cover the relevant predicates. Possible drawbacks include higher token usage and increased pressure on the model's context window, as the examples occupy a significant portion of the prompt.

*Chain-of-Thought (CoT)* explicitly instructs the model to reason about the prompt that was given and about the input log that was provided, before producing and outputting the structured RDF result. The LLM was instructed to have a hidden chain of thought and to perform the reasoning internally, since the rule of only outputting the RDF result must, still, be followed. In the prompt, it is explained to the LLM to decompose the main task into intermediate tasks such as tokenization, context extraction, deep scanning, validation, filtering, and even syntax check. Moreover, it also has some guidelines regarding the predicates, the KG model, the output constraints, and finally formatting rules to ensure that the output is a valid RDF (see Figure 4). One of the main benefits of using this method is that it is better at identifying the subject predicates and objects from complex loglines without having examples. However, for the small models, it can struggle to maintain a coherent logical chain of thought.

> [Internal Processing Protocol]
> 1. Tokenization: Split into Timestamp, PID, Level, Component, Message Body.
> 2. Context Extraction: Parse [ … ] block for Request ID, User ID, Tenant ID.
> 3. Deep Scanning: Identify UUIDs, IPs, numeric metrics, file paths, HTTP data.
> 4. Validation: Enforce correct datatype assignment.
> 5. Filtering: Only output predicates explicitly present. Never guess or infer.
> 6. Syntax Check: Ensure valid Turtle punctuation and termination.
> Never output this reasoning. Output only the RDF graph.
>
> [Ontology & Allowed Predicates]
> Core Metadata: log:logRecord, log:timestamp, log:processId, log:level, log:belongsToComponent…
> Context & Identity: log:requestId, log:belongsToUser, log:belongsToTenant
> Resource Links: log:belongsToInstance, log:hasImage, log:hasBaseFile, log:hasInstanceFile
> Network & HTTP: log:clientIp, log:serverIp, log:httpMethod, log:httpPath…
> Resource Metrics: Memory: …
>
> [Output Format Specification]
> Output ONLY the generated RDF in valid Turtle syntax with no hallucinations, markdown and prefixes

**Figure 4. CoT relevant components addition and updates**

*OSP & CoT* prompting is maintaining the hidden reasoning multi-step process presented before, while also giving the model the opportunity to see a real example to guide it for the result processing. In the prompt it is also specified or even enforced that the model should not rely on examples but process each log line that is presented individually. There are drawbacks associated such as single-point bias. Given the fact that the model is only presented with one example, it might not know how to reason and correlate with that example in the case of other various logs. There is also the token cost, since the prompt is becoming larger. There are, also, some advantages of the model, such as the demonstrated reasoning style, since the model is thought of how to perform, but also shown how to perform. Moreover, there is structure control, since the model has access to an example that is complete and correct.

*FSP & CoT prompting* is combining the CoT prompting method with multiple examples. In this case we have provided 27 examples (logs and RDF output result), just like in the Few-Shot, alongside with the chain of thought reasoning. The model is now exposed to multiple reasoning demonstrations, multiple examples, and it has the richest context of all the prompting techniques. However, this prompt is the longest from all the ones that were tested, which may be challenging with smaller models. This kind of combination (examples and reasoning) may have the highest accuracy for complex extraction tasks, such as the one that this project involves. This happens because the prompt can both teach the model the output format and the reasoning capabilities. Moreover, it is a robust method since the model has access to all the possible types of logs and all the possible output examples that are created through processing. There are also disadvantages associated with this technique. Sometimes the context saturation might happen, meaning that the model keeps reasoning while defocusing itself from the main task of extracting the semantic knowledge. Therefore, the model might produce comments, verbosity, even the reasoning process (even though explicitly told not to within the prompt) or repeat the examples. Also, there is the high token overhead. Since the prompt is quite long, the processing time increases and the cost, as well.

*Generate-Multiple-Then-Vote (GMV) prompting* is a strategy in which the model generates multiple candidate outputs and then selects the final output by voting for the best result. The prompt itself is rather simple, without examples (see Figure 5). In the beginning, the prompt is telling the model that the task of converting raw log lines into RDF triples using RDF structure needs to be performed. Then various guidelines are provided regarding the URIs, the predicates, the literals, and data types. In addition, there is the most important part, the multi-pass reasoning, where the model has multiple steps, including the Generate Candidate Internally step, the Voting Analysis Internally step, and the Final Output step. The first two steps are marked as internal to make sure that the model will not output that reasoning, or all the candidates, only the final one. The generation step has the purpose of generating three independently reasoned RDF outputs based on the instruction provided beforehand. The voting step is the one in which the model evaluates the three candidates and selects the best one. And then, the final step is when the final RDF block is marked as result and then outputted by the LLM. Besides these, there are also some formatting rules, specific to other prompts previously mentioned, as well. Also, the LLM is specifically told not to produce verbosity and to only output the RDF result. It is supposed to have higher accuracy since it's mitigating the hallucinations or the syntax errors that the LLMs might have in the context of semantic extraction. However, in this case, the limitations might be the large complexity, since the model is generating multiple results, and the higher cost.

```
[Internal Processing Protocol]
Before generating any output, you must perform the following reasoning steps internally only:
Step 1.  Candidate Generation: generate three independently reasoned RDF outputs.
Step 2. Self-Consistency Voting: compare the three generated candidates and select the most consistent and semantically correct one.
Step 3. Final Selection: Output only the selected RDF result.
You must not output the intermediate candidates or any reasoning process.
```
**Figure 5. Generate-Multiple-Then-Vote processing protocol**

*Self-Critique Prompting (SCP)* is a strategy in which the model is asked to generate the result and then to go back and revise it, critique it, and then refine it by making it better. This prompting technique is especially beneficial in identifying inconsistencies, missing predicates, in this case, missing properties, or even establishing if the model made certain mistakes during the reasoning process. The created prompt has basic instructions to convert the logs into RDF/Turtle output (see Figure 6). Then, the prompt is telling the model to perform all the self-critique reasoning internally. After, there are some guidelines or instructions regarding the name of the predicates, the URI rules, and so on. Finally, there is the most important part, the self-critique requirement section. Here, the model is instructed that before producing the final RDF output, it must silently verify if there are no invented fields, if all the fields that could have been extracted were included in the output, if the data types for the properties are correct, and if the turtle syntax is valid.

```
[Self-Critique Requirement]
Before producing the final RDF output, you must silently verify internally the following items:
     Ensure that only predicates explicitly present in the log line are included.
     All extractable fields from the log line are captured according to the ontology.
     All literals use the correct datatypes (xsd:dateTime, xsd:integer, xsd:decimal, or plain string).
     All URIs follow the prescribed patterns for components, users, tenants, instances, images, …
    The output strictly conforms to TTL syntax, including proper punctuation and termination rules.
     Each extracted value is correctly mapped to the corresponding semantic RDF predicate.
```
**Figure 6. Self-Critique Prompting relevant components added**

*Constraint Programming prompting (CPP)* defines the constraints that the model should follow rather than instructions that guide the reasoning process. The prompt starts with the normal task of transforming the log into RDF triples, followed by the constraints which are grouped into multiple sections (Figure 7). The first section is the input structure constraint, where the model is thought how the input will be provided to it. Then, we have the RDF output format constraint, where the model is told how an RDF or Turtle syntax must look like, followed by some strict rules regarding the syntax. After, the prompt has the allowed predicates that are used only if they are present in the logs (there are mandatory predicates that are present in all the logs, and then there are optional ones split into categories for a better understanding of them). A fourth constraint would be regarding the URI construction rules, where the model is taught how to construct the URIs that might appear in the result. Alongside this one, there are the types of rules for the literals. Then there is a non-hallucination rule, where the model is given the order to never guess, the component parsing rule, and then the absolute final output rule, where the model is told only to output the RDF without any comments. As disadvantages, this model might be prone to rigidity since the constraints are specific. For example, if there are properties that appear in the logs and are not present in the constraints, the model might not be able to tackle them.

> [Constraints]
> Rule 1: Process only the provided log line. Do not infer or invent values.
> Rule 2: Include only predicates present in the log line. Follow the ontology strictly.
> Rule 3: Construct URIs exactly as specified; never modify IDs.
> Rule 4: Use only valid datatypes: xsd:dateTime, xsd:integer, xsd:decimal, or plain string.
> Rule 5: Output RDF/Turtle. All lines except the last must end with ;. The last predicate must end with ..
> Rule 6: If a value cannot be extracted with 100% certainty you must omit it.
> Rule 7: Extract the component as the module path immediately after the log level

**Figure 7. Constraints as rules considered**

*Tree of Thought (ToT) prompting* strategy encourages the model to explore multiple alternative reasoning paths until choosing the right result. In our case the prompt contains the instruction of conversion and then an instruction that all the reasoning phases of the tree of thoughts process must remain internally, not present in the output. For creating this prompt, we have followed four specific stages (Figure 8). The first phase is a thought decomposition that is marked as internal and where the model is asked to generate at least three different interpretations of the log line and for each one to have a different viewpoint. The second phase is the branch evaluation, internally as well, when the model uses an internal evaluation matrix to calculate accuracy and other metrics to decide whether to keep that partial result, to prune it, or to refine it. The third phase is the refinement, where any promising interpretation up until that point must be re-evaluated internally and refined multiple times. The fourth phase is the selection, when the model is provided with some guidelines for selecting the right result, such as predicate examples, URI patterns, and others. Another extra phase is the final output format, when the model is thought how to output the actual result, what predicates to include, and what syntax to follow.

> [Internal Processing Protocol]
> Perform the following reasoning phases internally, without exposing them:
> PHASE 1: Thought Decomposition. Generate at least three independent interpretations of the log line.
> For each candidate, internally consider parsing approach, identified fields with confidence ranking, RDF structure.
> PHASE 2: Branch Evaluation. Score each candidate using completeness, accuracy, ontology compliance, URI consistency, and datatype correctness. Internally decide whether to keep, prune, or refine each candidate. Do not expose scores, tables, or analysis.
> PHASE 3: Refinement. Refine promising candidates and internally re-evaluate them. Do not describe refinement or evaluation.
> PHASE 4: Selection. Select the best candidate internally. Verify using this internal checklist: all mandatory fields present, no invented fields, all extractable conditional fields included, correct datatypes, Turtle syntax is valid, URIs follow prescribed patterns, predicates comply with the ontology. Output only the final RDF/Turtle triples.

**Figure 8. Tree of Thought internal processing protocol**

## 4. KG Construction and Reference Dataset

In this section, we present the design of the KG and its representation in RDF, leveraging cloud data logs and we introduce the reference Log-to-KG dataset, which serves as the basis for validating the RDF extraction and ensuring the consistency and quality of the generated knowledge graph.

### 4.1. KG RDF model

We have used the raw OpenStack cloud logs (around 200,000 lines) that are expressed as textual language, intended for human inspection, rather than automatic semantic processing. Even though the logs contain recurring metadata elements such as timestamps and IDs, they are not following a fix or explicit schema. Therefore, the logs have a semi-structured nature. Figure 9 shows an example from the data set emphasizing their heterogeneity, free-text variability, and contextual richness. Also, their structure is highly dependent on the context, some entries describing file system operations, others reporting HTTP requests and responses, while others capturing resource usage and scheduling decisions over the cloud infrastructure.

```
nova-api.log.1.2017-05-17_12:02:19
2017-05-16 18:57:49.073 25749 INFO nova.osapi_compute.wsgi.server
[req-0550be32-0499-40f3-b0cf-4aab2629052b
 113d3a99c3da401fbd62cc2caa5b96d254
 fadb412c4e40cdbaed9335e4c35a9e - - -]
10.11.10.1 "POST /v2/54fadb412c4e40cdbaed9335e4c35a9e/servers HTTP/1.1"
status: 202 len: 733 time: 0.4947891
...
nova-compute.log.2017-05-14_21:56:26
2017-05-14 21:30:50.117 2931 INFO nova.virt.libvirt.imagecache
[req-addc1839-2ed5-4778-b57e-5854eb7b8b09 - - - - -]
Removable base files:
/var/lib/nova/instances/_base/
a489c868f0c37da93b76227c91bb03908ac0e742
```

Figure 9. Cloud logs information example

Several recurring log characteristics have been identified and grouped into the semantic categories of entities described in Table 2. Not all attributes are present in every log entry, and each log type exposes a different subset of these properties. Therefore, we have designed a KG schema that formalizes core entities, their properties, and the relationships between them, ensuring consistency and enabling the accurate extraction of RDF triples from log data. The KG is intended to model entities along with their associated attributes. Relationships between these entities capture semantic links providing a structured representation of the underlying system behaviour. The graph itself is composed of nodes and edges, where the nodes represent subjects and objects, and the edges represent the predicates that connect them:

$$KG = (V, E) \qquad (1)$$

Where $V$ is a finite set of nodes representing subjects ($s$) and objects ($o$) and $E$ is a finite set of labelled edges representing predicates ($p$), with $P$ denoting the set of predicates. Each edge $(s, p, o) \in E$ encodes a semantic relationship between a subject $s \in V$ and an object $o \in V$ via a predicate $p \in P$.

The proposed KG adopts a log-centric conceptual model, in which each log entry represents a single semantic observation of system behaviour. Formally, let:

$$L = \{l_1, l_2, \ldots l_n\} \tag{2}$$

Each log entry $l_i \in L$ is mapped to a sub-graph $KG_i \subset KG$ such that:

$$KG = \cup_{i=1}^{n} KG_i \tag{3}$$

Each $KG_i$ represents the semantic interpretation of one log entry and contains one or more RDF triples derived from that entry.

**Table 2. Semantic categories on logs data**

| Entities | Attributes | Description |
|---|---|---|
| Component | belongsToComponent | Service or subsystem which generates the log entry |
| Log Metadata | Timestamp, Process ID, Log Level, Component ID, Request ID | Contextual information describing when, where, and by which component the log entry was generated |
| User and Tenant Context | User ID, Tenant ID | Identifies the user and project under which the operation was executed |
| Instance Context | Instance UUID | Unique identifier of the virtual machine involved in the operation |
| HTTP Interaction Information | Client IP, Server IP, HTTP Method, Request Path, Status Code, Response Length, Response Time | Network- and protocol-level details for request-driven operations |
| Image and File References | Image ID, Base File Path, Instance File Path | References to images and file system resources used during execution |
| Host Information | Host Identifier | Identifies the compute or controller node generating the log |
| Event Information | Event Type, Event ID, Event UUID | Describes the type and identity of system or operational events |
| Resource Usage | Memory, Disk, CPU | Reports resource availability, allocation, and utilization |
| Performance Metrics | Usage Counters, Timing Metrics, Status | Captures execution performance and resource claim outcomes |
| Free-Text Message Content | Log Message | Unstructured textual description of the operational event |

The KG is represented in RDF featuring triples are made in a structured way, representing a subject, a predicate, and an object:

$$KG = \{RDF_{triple} <s, p, o>, s \in V, o \in V, p \in P\} \tag{4}$$

where $s \in V$ is the set of RDF subjects, $o \in V$ is the set of RDF objects and $p \in P$ is the set of RDF predicates. We have defined rules for mapping and selecting each identified entities from the log lines to the elements of the RDF triple balancing expressiveness and simplicity. For subject ($s \in V$) identification, each log entry is represented by a unique RDF subject. The one-on-one mapping ensures that there is a direct traceability between the input data and the output RDF triple. Also, it ensures isolation of extraction and compatibility with the reference dataset that will be presented more in-depth in Section 4.2.

The predicates ($p \in P$) to capture the relationships that are commonly observed within the OpenStack cloud logs (see Table 3). Some of these predicates are related to: (i) contextual relationships such as association with components, user instances, or hosts, (ii) intrinsic attributes of the log entry such as timestamps or severity levels, and (iii) operational metrics and measurements such as computational resources usage. When the prompting engineering technique permitted, the predicate selection was restricted to a predefined set to prevent uncontrolled schema growth or even reduce semantic ambiguity.

**Table 3. The predicates that could have been extracted from the raw logs**

| Category | Predicate |
|---|---|
| **Metadata & Provenance** | Timestamp, ProcessId, Level, belongsToComponent, RequestId |
| **User, Tenant & Instance Context** | belongsToUser, belongsToTenant, belongsToInstance |
| **Network & HTTP Interaction** | ClientIp, ServerIp, HttpMethod, HttpPath, StatusCode, ResponseLength, ResponseTime |
| **Image, File & Host Information** | Image, BaseFile, InstanceFile, belongsToHost |
| **Event Information** | EventType, EventId, EventUuid |
| **Resource & Capacity Metrics** | PhysicalRamMB, UsedRamMB, TotalMemoryMB, UsedMemoryMB, FreeMemoryMB, hasMemoryLimitMB, PhysicalDiskGB, UsedDiskGB, TotalDiskGB, TotalVcpus, UsedVcpus, AllocatedVcpus, UsableVcpus, TotalUsableVcpus, TotalAllocatedVcpus |
| **Usage, Performance & Claim** | LocalUsageCount, RemoteUsageCount, SpawnTimeSeconds, BuildTimeSeconds, DurationSeconds, ClaimStatus, RequestedMemoryMB, RequestedDiskGB, RequestedVcpus, InstanceCount, UsageAuditStart, UsageAuditEnd |
| **Free-Text Content** | LogMessage |

The objects ($o \in V$) can be either resource or literal attributes. The resource objects are basically other entities that may be shared across multiple log entries. As an example, a user can be mentioned in multiple logs, just like a host, or an instance.

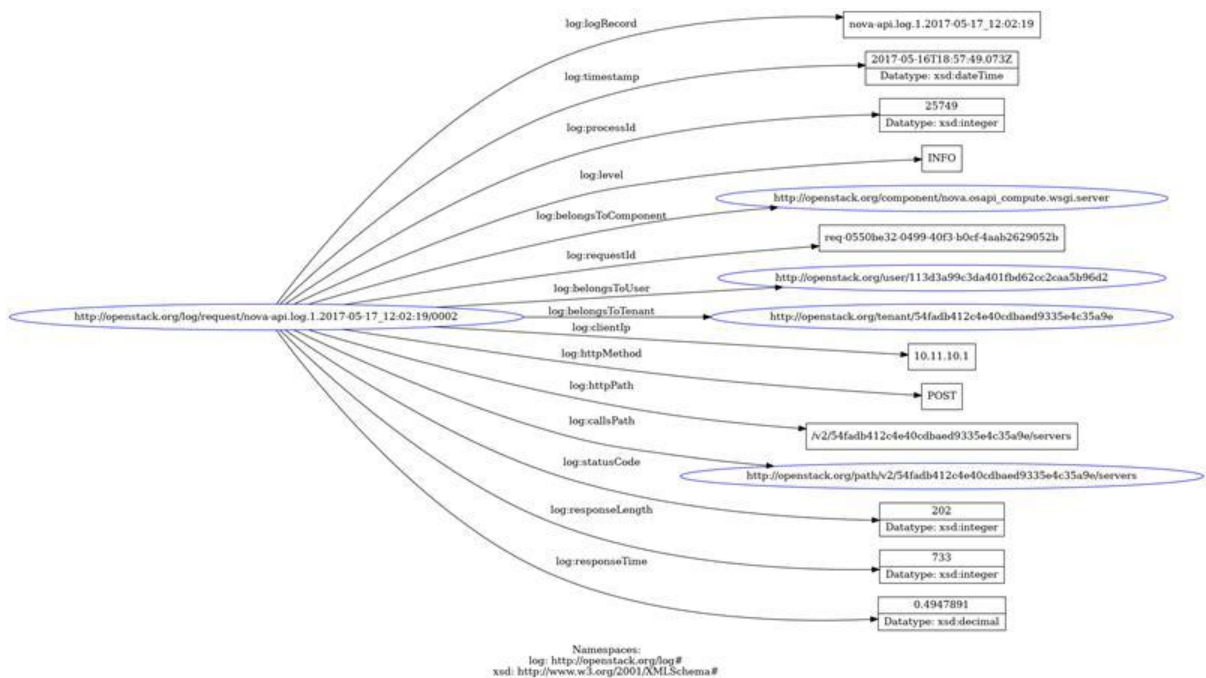

**Figure 10.** $KG_i$ **RDF example for a log entry**

By representing these elements as resources, the connectivity across the KG increases. The literal objects are portraying, mostly, numerical metrics, such as durations, timestamps, counters, or other measurements. The generation of the RDF triple is conditioned by the presence of the relevant information, so if any of the above elements are not present or cannot be reliably inferred from a log entry, the corresponding triple is omitted. Figure 10 shows an example of the $KG_i$ RDF representation of a log entry $i$ from the dataset.

## 4.2. Reference Log-to-KG dataset construction

The output generated by the LLMs can be non-deterministic and extremely sensitive to prompts, so there was a necessity for objective form of evaluation, since a manual inspection is a laborious work. Furthermore, LLMs may generate syntactically valid yet semantically incorrect triples that appear plausible at first glance. To address this, the evaluation must be performed at the level of individual log entries, using a reference dataset that reflects the single-entry processing paradigm through pairs of log entries and their corresponding RDF triples.

Since there is a lack of publicly available ground truth dataset for the evaluation of the LLM-based KG extraction we have created a reference Log-to-KG dataset leveraging OpenStack cloud logs [18]. The reference dataset creation is closely aligned with the RDF KG conceptual model, as each annotated dataset element represents a transformation example pairing a log entry with its corresponding RDF triple expressed in Turtle format:

$$f_{annotation}: l_i \rightarrow RDF_{triple} \qquad (5)$$

where is the annotation function used for constructing the reference dataset.

The Log-to-KG dataset contains 1,000 log entries selected from OpenStack log dataset each paired with its corresponding RDF triple. Some logs were similar but differed slightly in their attributes, and these variations were included to ensure diversity and maintain high-quality benchmarking. The types of logs encompass a range of operations, including API HTTP requests, metadata server events, instance lifecycle events, image cache management, resource claims and tracking, performance timing, instance file operations, event creation, usage audits, scheduling, and exception handling. Each log type contains both shared and unique properties, all of which are translated into RDF triples following the established schema.

For the construction of the reference Log-to-KG dataset we have used the following design guidelines to ensure high quality of the evaluation and the dataset itself. In the data set there should be an unambiguous and unique mapping between log entries and RDF triples. This one-on-one correspondence between the input and the output is essential for evaluation and benchmarking since it allows us to compare the automatically generated RDF response from the LLM to the one from the dataset. The naming of entities and predicates should be consistent throughout the reference dataset. Entities such as components, users, and tenants needed to have deterministic patterns, while the predicates came from predefined vocabulary. No other alternatives or synonyms were permitted during the annotation process since this design ensured that there were no semantic drift and reduced ambiguity during comparison. Finally, the RDF schema and RDF extraction rules had to be strongly related to how the reference Log-to-KG dataset was done.

The annotation process was done partially automatically and partially manually, meaning that after the automated annotation (log entries were transformed into RDF triples according to the defined modelling rules), each result was reviewed for consistency and semantic correctness (having human validation). In cases where the information was ambiguous, a conservative annotation strategy was adopted, and any semantic content that could not be represented in a structured form was intentionally omitted.

Figure 11 shows an example of annotated log entry in the dataset. The subject of the RDF is a URI that incorporates the log record, the timestamp, and a unique identifier. Next, the log contains a number (25749), which represents the process ID for the OpenStack operation. This is followed by the log level, indicated by the word "INFO," although other levels such as "CRITICAL" or "HIGH" could appear. The component that generated the log is then identified as "nova.osapi.compute.wsgi.server." Within square brackets, the log records the request ID, the user, and the tenant. After the brackets, a series of four numbers separated by dots appears, representing the client IP address. Finally, the log shows the actual request, starting with the HTTP method in this case, "POST" followed by the HTTP path, the response status (202), the request length (733), and the response time, which indicates how long it took to process the request.

```
nova -api.log .1.2017 -05 -17 _12 :02:19 2017 -05 -16 18:57:49.073 25749
INFO nova . osapi_compute . wsgi . server [req -0550 be32 -0499 -40 f3 -
b0cf -4 aab2629052b 113 d3a99c3da401fbd62cc2caa5b96d2 54
fadb412c4e40cdbaed9335e4c35a9e - - -] 10.11.10.1 " POST /v2 /54
fadb412c4e40cdbaed9335e4c35a9e / servers ␣ HTTP /1.1 " status : 202
len: 733 time : 0.4947891
```

```
<http://openstack.org/log/request/nova-api.log.1.2017-05-17_12:02:19/0002>
log:logRecord "nova-api.log.1.2017-05-17_12:02:19" ;
log:timestamp "2017-05-16T18:57:49.073Z"^^xsd:dateTime ;
log:processId "25749"^^xsd:integer ;
log:level "INFO" ;
log:belongsToComponent <http://openstack.org/component/nova.osapi_compute.
wsgi.server> ;
log:requestId "req-0550be32-0499-40f3-b0cf-4aab2629052b" ;
log:belongsToUser <http://openstack.org/user/113
d3a99c3da401fbd62cc2caa5b96d2> ;
log:belongsToTenant <http://openstack.org/tenant/54fadb412c4e40cdbaed9335e4c35a9e> ;
log:clientIp "10.11.10.1" ;
log:httpMethod "POST" ;
log:httpPath "/v2/54fadb412c4e40cdbaed9335e4c35a9e/servers" ;
log:callsPath <http://openstack.org/path/v2/54fadb412c4e40cdbaed9335e4c35a9e/servers> ;
log:statusCode "202"^^xsd:integer ;
log:responseLength "733"^^xsd:integer ;
log:responseTime "0.4947891"^^xsd:decimal .
```

**Figure 11. Annotated log line in the reference Log-to-KG dataset**

## 5. LLMs evaluation framework implementation

We have organized the implementation of the evaluation around two pipelines (see Figure 12): a LLM-based RDF extraction pipeline, and a RDF handling and validation pipeline. Each pipeline has distinct responsibilities and uses a dedicated set of tools and libraries, while together they support the end-to-end workflow of log transformation and model assessment.

LLM-based RDF extraction pipeline is implemented through a set of Python scripts, following a common structure, differing only in model-specific parameters and configurations. This pipeline consists of three stages. First, model loading and initialization are performed, where pretrained weights are loaded into memory using PyTorch [54], with automatic device placement on CPU or GPU and optional multi-GPU support via the Accelerate backend [55]. The selected LLM models employ a Transformer decoder architecture with causal attention and are executed using the Hugging Face Transformers framework [56]. Model-specific tokenizers are also initialized to ensure correct input encoding. Second, prompt construction and tokenization take place. Each prompt is assembled from a predefined template corresponding to a specific prompting strategy, combined with a single log entry and strict output-format instructions. The resulting prompt is then tokenized and transferred to the same device as the model to enable efficient inference. Finally, autoregressive generation is executed in inference mode, producing RDF triples in Turtle format [57]. Generation parameters are adjusted per model to control verbosity, determinism, and memory usage. Each log entry is processed independently, and the generated output is stored in model and prompting strategy associated result files.

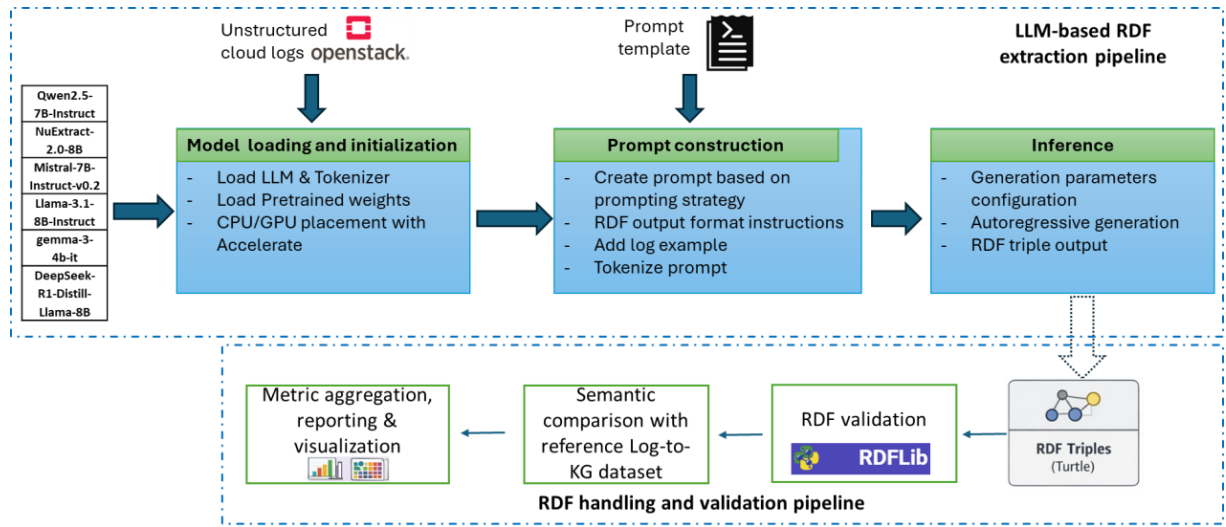

Figure 12. Evaluation framework pipelines

The second pipeline focuses on validation, comparison, and quantitative evaluation of the RDF triples generated by the LLMs. This pipeline is implemented as a multi-stage workflow operating on the outputs of the extraction pipeline. The first step performs RDF validation, where each generated result is parsed using the rdflib library [58]. Outputs are classified as fully valid RDF, partially valid (recoverable via regex-based extraction), invalid, or empty. This step ensures robustness against formatting errors and incomplete generations. Next, the evaluation workflow compares the validated triples against a manually constructed reference Log-to-KG dataset. The pipeline loads the triples from the Log-to-KG reference data set, iterates over all model–prompting combinations, and performs entry-level semantic comparison after normalization. Matched, missing, and hallucinated triples are identified, followed by predicate-level analysis to track confusion patterns and coverage. Finally, metric aggregation and reporting are performed. Precision, recall, F1-score, and predicate-specific metrics are computed, and visualized using matplotlib [59], through bar charts and heatmaps to support comparative analysis.

The OpenStack logs were processed using a single-entry approach. Each log line is taken and added to the prompt, then, the complete prompt is given to the LLM. Afterwards, the result is in form of RDF triples, ready to be used for automated KG construction. In this case, single-entry processing ensures strict traceability, establishing a one-to-one correspondence between each input log line and the generated RDF triples. This direct mapping is essential for interpreting model behaviour and validating the semantic correctness of the extracted knowledge. Also, when hallucinations or extraction errors occur, they can be traced back to a specific log entry, whereas in batch processing multiple outputs may be affected simultaneously. Additionally, this log processing method improves evaluation clarity, as it enables a direct and unambiguous comparison between the automatically generated triples and the corresponding entries in the manually constructed reference Log-to-KG dataset. Finaly, single-entry processing allows for better context control. Since LLMs operate within a finite context window and are sensitive to prompt structure and content, isolating each log entry ensures that the output is influenced solely by the input log and the intended prompt, without interference from surrounding data.

## 6. Evaluation results

We designed the evaluation process from two complementary perspectives: the structural syntactic reliability of RDF generation and the semantic extraction capabilities of the LLM models. The *syntactic evaluation* is done in a very strict manner, according to the reference Log-to-KG dataset. Each subject, predicate, and object must match the ones in the reference Log-to-KG dataset. This determines the ability of the model, combined with the prompting technique, to generate a fully schema-compliant RDF output. The *semantic evaluation* is more relaxed, allowing minor variation in formatting, in literal representation, or even in normalization of the identifiers. In this case, a generated triple is considered valid if the semantic relation present in the data was correctly captured, even if it was not fully compliant to the RDF output from the reference Log-to-KG dataset.

The LLM's output was cleaned by extracting only the RDF/TTL content when the models generated additional explanatory or verbose text and then was validated using a specific RDF library in Python. If that failed, we used a regex validation. As result the LLM output would fall into one of these categories: "Valid" if it was validated directly via the RDF library (and is a fully valid RDF result), "Regex" if it was validated via regex (but is not a fully valid RDF result), "Invalid" if it could have not been validated by any previous approach, or "Empty" if the LLM output was empty. Table 4 shows the RDF validation results across all six LLM models and ten prompting techniques. The RDF validation assesses the structural correctness of the generated RDF triples and determines whether they can be directly integrated into a knowledge graph generation pipeline without additional processing. These results reveal that the structural reliability of the RDF triples produced by the LLMs was heavily influenced not only by the prompt design, but also by the models. Also, it showcases the validity percentage that was calculated as a fraction between the fully valid results and the total number of results from the reference Log-to-KG dataset. It is important to note that the presence of invalid or partially valid RDF structures does not necessarily indicate the absence of meaningful semantic content.

**Table 4. RDF validation results per model and prompting technique**

| Prompting Technique | ZPT | ZPT & CoT | OSP | OSP & CoT | FSP | FSP & CoT | CCP | SCP | ToT | GMW |
|---|---|---|---|---|---|---|---|---|---|---|
| DeepSeek | | | | | | | | | | |
| Valid | 187 | 131 | 481 | 419 | 428 | 541 | 216 | 83 | 161 | 321 |
| Invalid | 476 | 18 | 25 | 6 | 25 | 1 | 10 | 2 | 41 | 19 |
| Empty | 70 | 0 | 1 | 1 | 0 | 1 | 1 | 0 | 17 | 0 |
| Regex | 267 | 851 | 493 | 574 | 547 | 457 | 773 | 915 | 781 | 660 |
| Valid (%) | 18.7 | 13.1 | 48.1 | 41.9 | 42.8 | 54.1 | 21.6 | 8.3 | 16.1 | 32.1 |
| gemma | | | | | | | | | | |
| Valid | 584 | 31 | 819 | 823 | 941 | 761 | 7 | 72 | 0 | 15 |
| Invalid | 284 | 913 | 70 | 0 | 0 | 0 | 30 | 1 | 26 | 38 |
| Empty | 0 | 0 | 80 | 0 | 40 | 238 | 527 | 473 | 470 | 333 |
| Regex | 132 | 56 | 31 | 177 | 19 | 1 | 436 | 454 | 504 | 614 |
| Valid (%) | 58.4 | 3.1 | 81.9 | 82.3 | 94.1 | 76.1 | 0.7 | 7.2 | 0.0 | 1.5 |
| Llama | | | | | | | | | | |
| Valid | 333 | 962 | 981 | 868 | 1000 | 1000 | 139 | 0 | 26 | 959 |
| Invalid | 299 | 0 | 0 | 0 | 0 | 0 | 6 | 0 | 10 | 2 |
| Empty | 0 | 0 | 0 | 0 | 0 | 0 | 6 | 0 | 0 | 0 |
| Regex | 368 | 38 | 19 | 132 | 0 | 0 | 849 | 1000 | 964 | 39 |

| Valid (%) | 33.3 | 96.2 | 98.1 | 86.8 | 100.0 | 100.0 | 13.9 | 0.0 | 2.6 | 95.9 |
|---|---|---|---|---|---|---|---|---|---|---|
| **Mistral** | | | | | | | | | | |
| Valid | 4 | 133 | 960 | 948 | 631 | 625 | 444 | 46 | 0 | 671 |
| Invalid | 207 | 25 | 0 | 0 | 0 | 0 | 0 | 827 | 41 | 78 |
| Empty | 95 | 0 | 0 | 0 | 0 | 34 | 0 | 0 | 0 | 0 |
| Regex | 694 | 842 | 40 | 52 | 369 | 341 | 556 | 127 | 959 | 251 |
| Valid (%) | 0.4 | 13.3 | 96.0 | 94.8 | 63.1 | 62.5 | 44.4 | 4.6 | 0.0 | 67.1 |
| **NuExtract** | | | | | | | | | | |
| Valid | 578 | 48 | 941 | 949 | 998 | 993 | 985 | 349 | 0 | 376 |
| Invalid | 418 | 0 | 0 | 0 | 0 | 0 | 0 | 0 | 40 | 480 |
| Empty | 0 | 0 | 0 | 0 | 1 | 0 | 0 | 0 | 375 | 21 |
| Regex | 4 | 952 | 59 | 51 | 1 | 7 | 15 | 651 | 585 | 123 |
| Valid (%) | 57.8 | 4.8 | 94.1 | 94.9 | 99.8 | 99.3 | 98.5 | 34.9 | 0.0 | 37.6 |
| **Qwen** | | | | | | | | | | |
| Valid | 6 | 912 | 1000 | 999 | 999 | 997 | 645 | 181 | 1 | 11 |
| Invalid | 0 | 38 | 0 | 0 | 0 | 0 | 0 | 118 | 153 | 93 |
| Empty | 0 | 0 | 0 | 0 | 0 | 0 | 0 | 0 | 0 | 0 |
| Regex | 994 | 50 | 0 | 1 | 1 | 3 | 355 | 701 | 846 | 896 |
| Valid (%) | 0.6 | 91.2 | 100.0 | 99.9 | 99.9 | 99.7 | 64.5 | 18.1 | 0.1 | 1.1 |

We can notice that there are certain best performance pairs of models and prompting strategies when it comes to the validity percentage. The combinations Llama with Few-Shot, Llama with Few-Shot and Chain-of-Thought, and Qwen with one-shot achieved 100% validity, while a NuExtract and Few-Shot technique reached 99.8% validity, which is near perfect with examples. This can be explained by the fact that, in the One-Shot and Few-Shot prompting techniques, the model is provided with one or more examples, which reinforces its ability to generate syntactically valid RDF triples when sufficient examples are available. Also, the worst performers combinations were Gemma with Tree-Of-Thought, Mistral with Tree-Of-Thought, Llama with Self-Critique, and Qwen with Zero-Shot. These produced the lowest valid RDF percentage across all the models, meaning that the complex reasoning chains might introduce errors when it comes to formatting or they might really deviate from the RDF syntax that was exposed or not in the prompt. In addition, as far as the prompting technique impacts, the best ones were Few-Shot and One-Shot, since providing examples dramatically improves RDF validity, while without examples, in cases of Zero-Shot, Tree-of-Thought, or Self-Critique, the models really struggled to produce the correct syntax.

For the models and prompting techniques that exhibited a high reliance on regex-based fallback analysis, the outputs were largely similar to valid RDF but contained multiple syntactic errors, such as missing semicolons, incorrect prefix declarations, or malformed URIs. An interesting finding from this validation evaluation is the combination of Llama with the Self-Critique technique, which resulted in 0% fully valid RDF triples. However, all outputs could be processed using regex-based validation. This indicates that, although the generated triples closely followed the RDF schema in terms of structure, they contained syntactic errors, such as missing semicolons, incorrect prefixes, or malformed URIs, that prevented strict RDF validation. The regex approach was able to identify and extract these near-RDF patterns, showing that the outputs still contained meaningful structural information, even if they were not formally correct. Also, there were many empty outputs for Gemma with either Constraint Prompting, Self-Critique, or even Tree-Of-Thought, where the model simply failed to generate the RDF.

Finally, for metrics calculation each subject, predicate and object were taken separately and compared with the ones in the reference Log-to-KG dataset for the specific log-RDF pair. Each predicate and object was systematically examined to evaluate potential hallucinations by the LLM, verify presence in the reference Log-to-KG dataset, and ensure syntactic correctness and completeness. Several metrics were employed for the evaluation of LLM-based RDF extraction, each reflecting different aspects of performance. *Precision* was used to measure the proportion of RDF triples generated that are correct compared to the reference Log-to-KG dataset. It is important because hallucinated RDF triples can introduce misleading information into a supposedly constructed KG, which would make the debugging and the understanding of the system that produced the logs harder. *Recall* determines the proportion of reference RDF triples that are successfully extracted by each model. The recall is a metric just as important as precision, because the OpenStack logs might contain implicit information that could not be extracted or transformed into predicates, such as resource usage or instance association, things that are present mostly in the message part of the log. A high value of recall means that the extracted graph captures all the information presented in the logs. *F1-score* determines the harmonic mean of precision and recall. The F1 score represents the trade-off between precision and recall, and it is a very powerful metric when evaluating LLMs. Sometimes, prompting strategies would increase the recall at the risk of hallucinations or would simply favour precision. However, a balance needs to be established between them. Finally, *Micro F1-score* is particularly well-suited due to the highly imbalanced distribution of the predicates across the logs. They tend to contain a small number of predicates that are occurring all the time, alongside many infrequent ones or context-dependent predicates, considering the type of the log. Hence, the Micro F1 aims to capture the balance between precision and recall at a global level, reflecting both the correctness of the RDF triples and the completeness of the semantic information that was extracted from the logs.

### 6.1. Syntactic RDF evaluation

The syntactic evaluation is meant to showcase which model and prompting technique combination can provide the best results in terms of correct syntax of KG RDF triples extracted from logs. Table 5 presents detailed values for precision, recall, and F1 score across all the models and all the prompting engineering strategies that were compared. These results show that there is big variability across all models and prompting techniques, and this indicates that the extraction performance does not depend on the LLM alone. By far, the best prompting technique was the one involving Few-Shot, since it achieved the highest micro F1 score, and, in several combinations with different models, it was approaching a near-perfect performance (see also figure 13). This strongly emphasizes the fact that when the model receives multiple examples of structured output, it becomes highly effective in semantic extraction. Even the combination of Few-Shot and Chain-of-Thought outperformed other types of prompting, particularly in recall. The Zero-Shot prompting had a very low recall across all models, resulting, then, in a poor F1 score, despite having moderate precision sometimes. This means that without explicit examples, the model struggles to produce results that are syntactically correct and according to the ones from the reference Log-to-KG dataset. Even when the Chain-of-Thought characteristic was utilized to enhance the Zero-Shot prompting technique, the improvements were not quite significant. The Constraint Prompting technique improved structural correctness when combined with certain models, but, also, had lower recall and lower F1 scores, meaning that all the constraints given to the model might have limited the ability to extract the semantic relations that were not very explicit. Strategies such as Self-Critique, Tree-Of-Thought, and Generate Multiple are more advanced ones, but the models were not behaving well using these strategies, resulting in low precision and recall. In some evaluation cases the model would not produce any triples since the prompt was, most probably, confusing for the output generation. Regarding the model behaviour, that can be deduced from this table, we can state that instruction-tuned models, such as Llama, Qwen, or NuExtract, achieved high precision and recall with Few-Shot prompting, since this aligns with their training and their extraction-tuned characteristics. At the same time, smaller models or even less specialized ones (in extracting structured information) become more sensitive to different prompts or to prompts that require an explicit critical thinking chain.

**Table 5. syntactic evaluation results across models and prompting techniques**

| Prompting Technique | ZPT | ZPT & CoT | OSP | OSP & CoT | FSP | FSP & CoT | CCP | SCP | ToT | GMW |
|---|---|---|---|---|---|---|---|---|---|---|
| **DeepSeek** | | | | | | | | | | |
| Precision | 0.0822 | 0.0864 | 0.4550 | 0.4227 | 0.4339 | 0.5309 | 0.1573 | 0.0511 | 0.1138 | 0.1967 |
| Recall | 0.0215 | 0.0765 | 0.4061 | 0.3875 | 0.3972 | 0.5057 | 0.1140 | 0.0420 | 0.0901 | 0.1784 |
| F1 | 0.0341 | 0.0811 | 0.4291 | 0.4044 | 0.4147 | 0.5180 | 0.1322 | 0.0461 | 0.1005 | 0.1871 |
| **Gemma** | | | | | | | | | | |
| Precision | 0.2307 | 0.2824 | 0.6871 | 0.8202 | 0.9662 | 0.9505 | 0.0050 | 0.0814 | 0.0000 | 0.0133 |
| Recall | 0.1413 | 0.0181 | 0.7236 | 0.8108 | 0.8955 | 0.7599 | 0.0014 | 0.0216 | 0.0000 | 0.0062 |
| F1 | 0.1752 | 0.0340 | 0.7049 | 0.8154 | 0.9295 | 0.8445 | 0.0022 | 0.0341 | 0.0000 | 0.0084 |
| **Llama** | | | | | | | | | | |
| Precision | 0.1281 | 0.6640 | 0.6682 | 0.7841 | 0.9897 | 0.9921 | 0.0644 | 0.0000 | 0.0079 | 0.4690 |
| Recall | 0.0910 | 0.7409 | 0.8677 | 0.7984 | 0.9974 | 0.9944 | 0.0690 | 0.0000 | 0.0097 | 0.7343 |
| F1 | 0.1064 | 0.7003 | 0.7550 | 0.7912 | 0.9935 | 0.9932 | 0.0667 | 0.0000 | 0.0087 | 0.5724 |
| **Mistral** | | | | | | | | | | |
| Precision | 0.0012 | 0.0488 | 0.8317 | 0.8491 | 0.5302 | 0.5605 | 0.1890 | 0.1466 | 0.0000 | 0.3004 |
| Recall | 0.0007 | 0.0351 | 0.8477 | 0.7944 | 0.5937 | 0.5719 | 0.1811 | 0.0383 | 0.0000 | 0.5488 |
| F1 | 0.0009 | 0.0408 | 0.8396 | 0.8209 | 0.5602 | 0.5662 | 0.1850 | 0.0607 | 0.0000 | 0.3882 |
| **NuExtract** | | | | | | | | | | |
| Precision | 0.1754 | 0.0349 | 0.6740 | 0.8760 | 0.9823 | 0.9846 | 0.8071 | 0.2674 | 0.0000 | 0.3524 |
| Recall | 0.0495 | 0.0296 | 0.8330 | 0.8758 | 0.9860 | 0.9775 | 0.7039 | 0.2162 | 0.0000 | 0.2043 |
| F1 | 0.0772 | 0.0320 | 0.7451 | 0.8759 | 0.9841 | 0.9810 | 0.7520 | 0.2391 | 0.0000 | 0.2586 |
| **Qwen** | | | | | | | | | | |
| Precision | 0.0013 | 0.6030 | 0.8162 | 0.9596 | 0.9926 | 0.9872 | 0.5612 | 0.1162 | 0.0000 | 0.0081 |
| Recall | 0.0011 | 0.5122 | 0.8773 | 0.9039 | 0.9906 | 0.9883 | 0.4296 | 0.0810 | 0.0000 | 0.0089 |
| F1 | 0.0012 | 0.5539 | 0.8456 | 0.9309 | 0.9916 | 0.9878 | 0.4866 | 0.0954 | 0.0000 | 0.0085 |

Several combinations of prompting technique and LLM had high precision but lower recall, indicating conservative extraction behaviour that would prioritize the correctness of the result over the completeness.

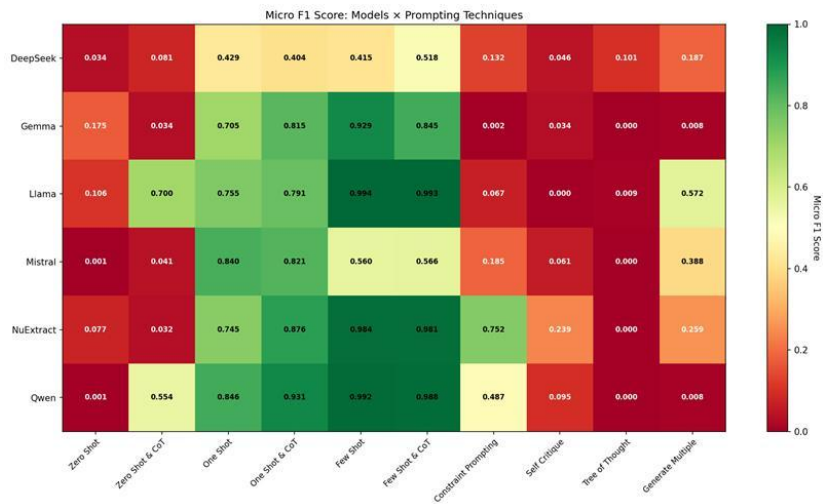

**Figure 13. F1 Heatmap for syntactic evaluation**

In consequence, some prompting strategies would improve recall with the cost of precision, reflecting a tendency to over-generate multiple RDF triples. The micro F1 score that was computed during the evaluation showcases this trade-off and highlights the configurations that achieved a balanced extraction behaviour in the case of syntactic evaluation. This can be also observed in Figure 13, where is the F1 heatmap with the micro F1 score for all the models and prompting techniques. As shown in Figure 14, for syntactic evaluation, the prompting techniques that performed best were Few-Shot and One-Shot with Chain-of-Thought. These techniques achieved substantially higher F1 scores than the other six prompting strategies, indicating their superior ability to generate syntactically correct outputs.

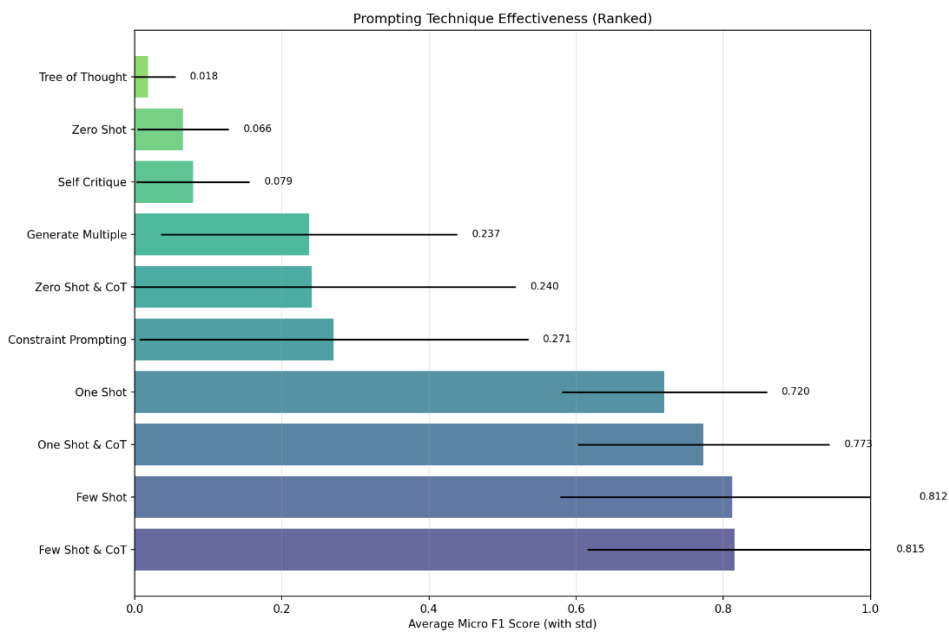

Figure 14. Prompting technique effectiveness in syntactic evaluation

## 6.2. Semantic RDF Evaluation

The semantic evaluation is more relaxed, and it focuses on the ability of the models, combined with the prompting technique, to extract the semantic information, even though it is not compliant to the schema from the reference Log-to-KG dataset. For example, if in the reference Log-to-KG dataset the predicate is "httpPath", in the result it would be completely valid as "HTTP_PATH".

Table 6 presents the results for precision, recall and F1 score in case of semantic evaluation, where minor deviations in RDF formatting or literal representation are tolerated, if the underlying semantic relation is correctly identified. In this case for each combination of LLM and prompting technique, the results are substantially higher since the evaluation is more semantically focused, rather than syntactically. The Few-Shot prompting technique achieved the highest semantic extraction performance for all evaluated models, even having micro F1 scores above 0.98 with multiple configurations. This emphasizes, once again, that the LLM models behave better if they are provided with examples. In addition, even in combination with Chain-of-Thought, the Few-Shot technique outperformed the other methods. In the case of One-Shot prompting, it has been a moderate semantic performance, while Zero-Shot remained less effective. This shows that even by providing only one example to the model, the precision and recall increase significantly. However, by adding Chain-of-Thought to this technique the results would not improve significantly. In this case of evaluation, Constraint Prompting improves precision, but is quite limited in recall, proving that all the structural constraints are restricting the extraction. The advanced prompting strategies, such as Self-Critique or Tree-Of-Thought often failed to identify relevant relations, suggesting that an increased reasoning process does not necessarily mean that the semantic extraction task will have better results.

**Table 6. Detailed semantic evaluation results across models and prompting techniques**

| Prompting Technique | ZPT | ZPT & CoT | OSP | OSP & CoT | FSP | FSP & CoT | CCP | SCP | ToT | GMW |
|---|---|---|---|---|---|---|---|---|---|---|
| DeepSeek | | | | | | | | | | |
| Precision | 0.2209 | 0.7270 | 0.8212 | 0.8836 | 0.9227 | 0.9388 | 0.7258 | 0.7139 | 0.6681 | 0.6101 |
| Recall | 0.0577 | 0.6435 | 0.7328 | 0.8099 | 0.8447 | 0.8943 | 0.5262 | 0.5862 | 0.5286 | 0.5533 |
| F1 | 0.0915 | 0.6827 | 0.7745 | 0.8451 | 0.8820 | 0.9160 | 0.6101 | 0.6438 | 0.5902 | 0.5803 |
| Gemma | | | | | | | | | | |
| Precision | 0.2947 | 0.5430 | 0.7033 | 0.9141 | 0.9855 | 0.9516 | 0.4076 | 0.4644 | 0.3507 | 0.3845 |
| Recall | 0.1805 | 0.0347 | 0.7407 | 0.9036 | 0.9134 | 0.7608 | 0.1134 | 0.1232 | 0.0785 | 0.1778 |
| F1 | 0.2239 | 0.0653 | 0.7215 | 0.9088 | 0.9481 | 0.8456 | 0.1774 | 0.1948 | 0.1283 | 0.2432 |
| Llama | | | | | | | | | | |
| Precision | 0.2610 | 0.6821 | 0.6752 | 0.8485 | 0.9897 | 0.9921 | 0.6116 | 0.4149 | 0.4915 | 0.4865 |
| Recall | 0.1851 | 0.7612 | 0.8768 | 0.8640 | 0.9974 | 0.9944 | 0.6553 | 0.6094 | 0.6012 | 0.7617 |
| F1 | 0.2166 | 0.7195 | 0.7629 | 0.8562 | 0.9935 | 0.9932 | 0.6327 | 0.4937 | 0.5408 | 0.5937 |
| Mistral | | | | | | | | | | |
| Precision | 0.2680 | 0.6600 | 0.8572 | 0.8993 | 0.8534 | 0.9231 | 0.5533 | 0.5098 | 0.3784 | 0.3972 |
| Recall | 0.1442 | 0.4741 | 0.8735 | 0.8413 | 0.9551 | 0.9419 | 0.5304 | 0.1330 | 0.4103 | 0.7256 |
| F1 | 0.1875 | 0.5518 | 0.8653 | 0.8693 | 0.9014 | 0.9324 | 0.5416 | 0.2110 | 0.3937 | 0.5134 |
| NuExtract | | | | | | | | | | |
| Precision | 0.1766 | 0.5852 | 0.7057 | 0.9047 | 0.9829 | 0.9905 | 0.8169 | 0.6312 | 0.4651 | 0.4922 |
| Recall | 0.0498 | 0.4952 | 0.8722 | 0.9044 | 0.9865 | 0.9833 | 0.7125 | 0.5103 | 0.2983 | 0.2853 |
| F1 | 0.0777 | 0.5365 | 0.7801 | 0.9045 | 0.9847 | 0.9869 | 0.7611 | 0.5644 | 0.3635 | 0.3612 |
| Qwen | | | | | | | | | | |
| Precision | 0.3541 | 0.7536 | 0.8162 | 0.9604 | 0.9932 | 0.9902 | 0.8441 | 0.7178 | 0.6630 | 0.6523 |
| Recall | 0.2967 | 0.6402 | 0.8773 | 0.9046 | 0.9912 | 0.9913 | 0.6461 | 0.5002 | 0.2545 | 0.7145 |
| F1 | 0.3229 | 0.6923 | 0.8456 | 0.9317 | 0.9922 | 0.9907 | 0.7320 | 0.5895 | 0.3678 | 0.6820 |

Figure 15 presents the F1 heatmap for semantic evaluation. The F1 scores increased overall since both precision and recall values increased, because of the relaxed manner of the evaluation.

Figure 16 presents the prompting technique effectiveness of the semantic evaluation where the prompting strategies are ranked by the average micro F1 score. Like before, the best ones were Few-Shot, Few-Shot with Chain-of-Thought, One-Shot with Chain-of-Thought, One-Shot, and then the others. The worst one was, obviously, Zero-Shot, since the model did not provide any examples, instructions related to the schema or the predicates, or even how it should think during the extraction. As a comparison between the syntactic and semantic evaluation, the main discrepancy was noticed for predicates involving numerical values, resource usage, or contextual identifiers, since they were often semantically correct, but they would be invalid due to formatting.

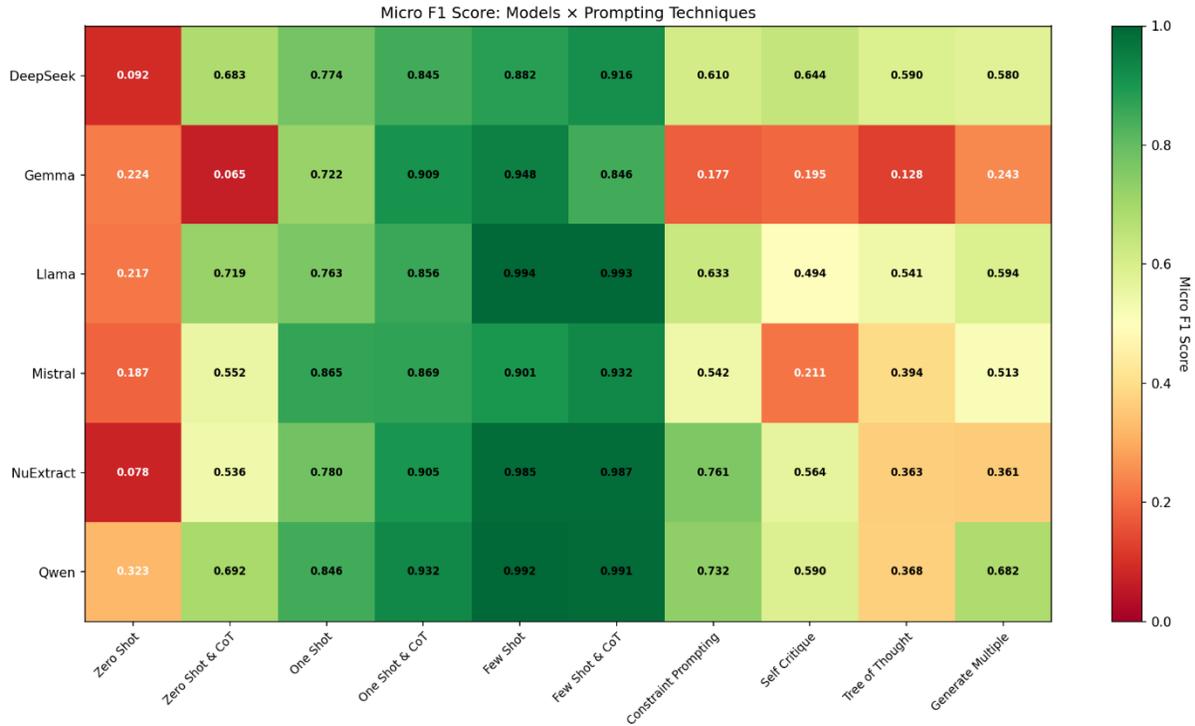

**Figure 15. F1 Heatmap for semantic evaluation**

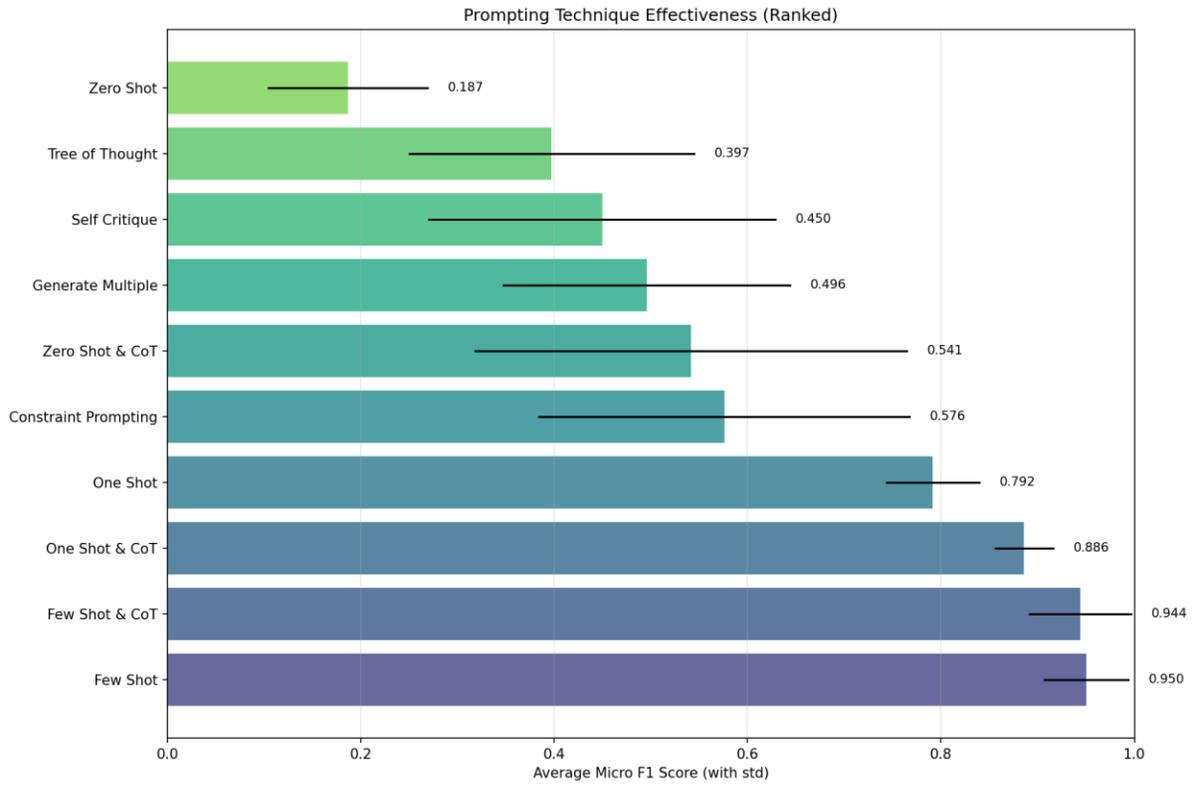

**Figure 16. Prompting technique effectiveness for semantic evaluation**

## 7. Discussion

In this section we discuss two important aspects of the RDF extraction evaluation. First, we analyse the results at the predicate level, highlighting the predicates most frequently missed by each prompting technique and their occurrence rates, as well as the most missed and hallucinated predicates across all models, considering the semantic evaluation. We also examine the most confused predicate pairs, providing insights into patterns of misclassification and ambiguity. Second, each model and prompting technique is ranked using the micro F1 score to assess overall performance under both syntactic and semantic evaluation criteria, allowing a comparative understanding of which configurations perform best in generating accurate and meaningful RDF triples.

Table 7 shows the most missed predicates (in syntactic evaluation) are also the ones that tend to appear in all or in most of the logs, including *logRecord*, *level*, *processId*, *requestId*, *timestamp*, or *belongsToComponent*. As far as the missed predicates, they tend to be missed in techniques such as Zero-Shot or One-Shot, where the model is not provided with examples or is provided only one example. In addition, there are also a lot of missed predicates in advanced techniques, such as Self-Critique, Tree-of-Thought, and Generate Multiple.

**Table 7. Top 5 missed predicates categorized by prompting technique (syntactic evaluation)**

| Technique | Missed Predicate | Frequency | Technique | Missed Predicate | Frequency |
|---|---|---|---|---|---|
| ZSP | logRecord | 6000 | FSP & CoT | timestamp | 1130 |
| | timestamp | 5951 | | logRecord | 1090 |
| | processId | 5680 | | belongsToComponent | 1088 |
| | belongsToComponent | 5678 | | level | 1083 |
| | level | 5065 | | processId | 1083 |
| ZSP & CoT | logRecord | 5546 | CPP | timestamp | 5582 |
| | belongsToComponent | 5306 | | logRecord | 4538 |
| | timestamp | 5003 | | requestId | 4238 |
| | processId | 4198 | | processId | 3911 |
| | requestId | 4174 | | belongsToComponent | 3871 |
| OSP | message | 2613 | SCP | timestamp | 5712 |
| | belongsToInstance | 1544 | | belongsToComponent | 5660 |
| | logRecord | 1090 | | logRecord | 5523 |
| | hasBaseFile | 1014 | | requestId | 5291 |
| | timestamp | 843 | | level | 5290 |
| OSP | message | 2115 | ToT | logRecord | 6000 |
| | callsPath | 1739 | | timestamp | 5984 |
| | requestId | 1143 | | belongsToComponent | 5895 |
| | belongsToComponent | 1065 | | processId | 5825 |
| | timestamp | 1012 | | level | 5814 |
| Few-Shot | logRecord | 1015 | GMV | logRecord | 5442 |
| | belongsToComponent | 1008 | | timestamp | 5089 |
| | timestamp | 1004 | | belongsToComponent | 4154 |
| | level | 1004 | | processId | 3757 |
| | processId | 1004 | | requestId | 3746 |

Table 8 summarizes the most missed predicates (false negatives) and the most hallucinated ones (false positives), across all models and prompting techniques under semantic evaluation. Some of the most missed predicates are *timestamp*, *logRecord*, *callsPath*, *requestId*, *message*, or level, predicates that tend to appear the most in the OpenStack logs. These are missing because they are often appearing implicitly, not explicitly, they are embedded in long prefixes, or absent from the message body. Fields like message, *clientIp*, *serverIp*, *httpPath*, and *httpMethod* are more prone to hallucination because they are often open-ended or highly variable, under-represented in training data, and sometimes lack clear format constraints. LLMs therefore generate plausible but inaccurate values, especially when context is limited or no strict schema is provided. They tend to be even clearer One-Shot results, since the one example that is given is an HTTP request log.

**Table 8. Top 10 most missed and hallucinated predicates across all models (semantic evaluation)**

| Rank | Most Missed (False Negative) | Most Hallucinated (False Positive) |
|---|---|---|
| 1 | timestamp | message |
| 2 | logRecord | belongsToInstance |
| 3 | belongsToComponent | belongsToTenant |
| 4 | callsPath | belongsToUser |
| 5 | requestId | clientIp |
| 6 | belongsToTenant | serverIp |
| 7 | belongsToUser | httpPath |
| 8 | message | httpMethod |
| 9 | processId | statusCode |
| 10 | level | responseTime |

Table 9 presents the top 10 pairs of predicates that were confused during the RDF extraction. Predicates such as *belongsToInstance* was confused with the *instanceId*, the *belongsToComponent* was confused with the *logger*, the *statusCode* was perceived as *status*, *belongsToInstance* was perceived as *Instance*, and so on. As can be seen from the table, in most cases, the predicate was just a shortened version, semantically correct, of the predicate version from the reference Log-to-KG dataset. This is because they prioritize semantic plausibility over exact schema names. Fields with similar meaning, under-specified prompts, or common textual patterns are easily conflated, leading to simplified or misinterpreted labels. This also shows the ability of LLMs to extract meaningful semantic information from unstructured OpenStack log entries and to infer relevant contextual relations.

**Table 9. Top 10 predicate confusion pairs (A omitted → B produced)**

| Reference Predicate (A) | Confused Predicate (B) | Frequency |
|---|---|---|
| belongsToInstance | instanceId | 1,291 |
| belongsToComponent | logger | 1,126 |
| statusCode | status | 989 |
| belongsToInstance | instance | 707 |
| httpMethod | requestMethod | 577 |
| responseLength | length | 529 |
| hasBaseFile | hasInstanceFile | 503 |
| requestId | logRecord | 470 |
| belongsToUser | belongsToTenant | 456 |
| httpMethod | method | 407 |

In terms of model and prompting technique performance we ranked them by the micro F1 score to see how they performed the best, under both syntactic and semantic evaluation. Table 10 presents the best performing model for each prompting technique under syntactic evaluation. Gemma was the best model in combination with Zero-Shot, Llama was the best model in combination with Zero-Shot with Chain-of-Thought, Few-Shot, Few-Shot with Chain- of-Thought, and Generate Multiple. In addition, Qwen was the best model for both One-Shot and One-Shot with Chain-of-Thought. NuExtract was the best for Constraint Prompting and Self-Critique, while DeepSeek proved to be the best for Tree-Of-Thought.

**Table 10. Best performing model for each prompting technique based on Micro F1 score (syntactic evaluation)**

| Technique | Best Model | Micro F1 |
|---|---|---|
| ZSP | Gemma | 0.1752 |
| ZSP & CoT | Llama | 0.7003 |
| OSP | Qwen | 0.8456 |
| OSP & CoT | Qwen | 0.9309 |
| FSP | Llama | 0.9935 |
| FSP & CoT | Llama | 0.9932 |
| CPP | NuExtract | 0.7520 |
| SCP | NuExtract | 0.2391 |
| ToT | DeepSeek | 0.1005 |
| GMV | Llama | 0.5724 |

In table 11 we present the best performing prompting technique for each model based on micro F1 score, under syntactic evaluation. The Few-Shot prompting technique was the best one in combination with Gemma, Llama, NuExtract, and Qwen, while Few-Shot with Chain-of-Thought was the best for DeepSeek. In addition, Mistral and One-Shot achieved 0.83 in micro F1 score.

**Table 11. Best performing prompting technique for each model based on Micro F1 score (syntactic evaluation)**

| Model | Technique | Micro F1 |
|---|---|---|
| DeepSeek | FSP & CoT | 0.5180 |
| Gemma | FSP | 0.9295 |
| Llama | FSP | 0.9935 |
| Mistral | OSP | 0.8396 |
| NuExtract | FSP | 0.9841 |
| Qwen | FSP | 0.9916 |

Table 12 presents the best performing model for each prompting technique, considering semantic evaluation. In this case, techniques such as Zero-Shot, One-Shot with Chain-of-Thought, and Generate Multiple achieved the best results with the model Qwen. Zero-shot with Chain-of-Thought, Few-Shot, and Few-Shot with Chain-of-Thought had the best micro F1 in combination with Llama. One-Shot prompting technique had the best result with Mistral, Constraint Prompting with NuExtract and the advanced prompting techniques Self-Critique and Tree-Of-Thought had the best results with DeepSeek.

**Table 12. Best performing model for each prompting technique based on Micro F1 score (semantic evaluation)**

| Technique | Best Model | Micro F1 |
|---|---|---|
| ZSP | Qwen | 0.3229 |
| ZSP & CoT | Llama | 0.7195 |
| OSP | Mistral | 0.8653 |
| OSP & CoT | Qwen | 0.9317 |

| FSP | Llama | 0.9935 |
| FSP & CoT | Llama | 0.9932 |
| CPP | NuExtract | 0.7611 |
| SCP | DeepSeek | 0.6438 |
| ToT | DeepSeek | 0.5902 |
| GMV | Qwen | 0.6820 |

Table 13 presents the best performing prompting techniques for each specific model under semantic evaluation. For half of the models, the best prompting technique proved to be Few-Shot, and for the other half, Few-Shot with Chain-of-Thought. In all these cases, the micro F1 score reached results above 0.91, meaning that the LLMs performed the best when they were given multiple examples and they were told how to establish the thinking process for the knowledge extraction along with the examples.

Table 13. Best performing prompting technique for each specific model based on Micro F1 score (semantic evaluation)

| Model | Technique | Micro F1 |
|---|---|---|
| DeepSeek | FSP & CoT | 0.9160 |
| Gemma | FSP | 0.9481 |
| Llama | FSP | 0.9935 |
| Mistral | FSP & CoT | 0.9324 |
| NuExtract | FSP & CoT | 0.9869 |
| Qwen | FSP | 0.9922 |

## 8. Conclusions

In this paper, we evaluated the effectiveness and efficiency of various LLMs and fine-tuning strategies for automated extraction of structured RDF knowledge graphs from cloud system logs. To enable a systematic and comparative assessment, we developed a framework with two pipelines that provide a controlled testbed for applying selected LLMs to heterogeneous log data. This setup allowed us to analyse LLMs performance across different prompt engineering strategies and to identify the strengths and limitations of each approach. We have provided detailed information on the models' selection, technology stack and reference Log-to-KG construction used in results validation, to ensure reproducibility and transparency of our experiments.

The experimental results show that the LLMs performance in extracting KG RDF triples from unstructured logs, is highly dependent on the prompting strategy. Few-Shot prompting strategy consistently achieved the highest F1 scores and very good RDF validity across all evaluated models. In this context very good accuracy was provided by Llama, Qwen, NuExtract, and Gemma. One-Shot prompting provided a lighter but still effective alternative, achieving good performance above 85% F1 score. In contrast, Zero-Shot approaches largely failed, producing low F1 scores and minimal valid outputs. More advanced prompting techniques such as Tree-of-Thought, Self-Critique, and Generate-Multiple improved performance only partially and could not match Few-Shot baselines. These results highlight the importance of contextual examples, prompt design and model architecture for reliable RDF extraction.

As future work, we plan to focus on autonomous knowledge graph exploration, enabling advanced querying, reasoning, and inference over the extracted RDF knowledge graphs. Specifically, we aim to investigate how LLMs can facilitate intelligent navigation, contextual query generation, and automated reasoning within these KG, closing the gap between raw log data and actionable insights. We also intend to explore hybrid approaches combining symbolic reasoning with LLM capabilities and evaluate their performance in enhancing accuracy, and interpretability in autonomous graph exploration.

**Acknowledgement**

This work was supported by the project "Romanian Hub for Artificial Intelligence-HRIA", Smart Growth, Digitization and Financial Instruments Program, MySMIS no. 334906.